\documentclass{elsart}
\usepackage{graphicx}

\def\textbf#1{{\bf #1}}
\def\textit#1{{\it #1}}

\begin{document}

% ------------------------------
\begin{frontmatter}

\title{Power law for ensembles of stock prices}
\author[Tokyo1,Tokyo2]{Taisei Kaizoji\thanksref{contact}}
\author[Tokyo2]{Michiyo Kaizoji}
\address[Tokyo1]{%
  Division of Social Sciences,
  International Christian University, Osawa, Mitaka, Tokyo 181-8585,
  Japan}
\address[Tokyo2]{%
  Econophysics Laboratory, 5-9-7-B Higashi-cho, Koganei-shi, Tokyo, 184-0011, Japan.}
\thanks[contact]{%
Corresponding author. E-mail: kaizoji@icu.ac.jp, \\
URL: http://members.jcom.home.ne.jp/ephys/e-index.html}

\begin{abstract}
In this paper we quantitatively investigate the statistical properties of an ensemble of {\it stock prices}. We selected $1200$ stocks traded in the Tokyo Stock Exchange and formed a statistical ensemble of daily stock prices for each trading day in the 5 year period from January 4, 1988 to December 30, 1992. We found that the tail of the complementary cumulative distribution function of the ensembles is accurately described by a power-law distribution with an exponent that moves in the range of $ 1.7 < \alpha < 2.2 $. 
\end{abstract}

\end{frontmatter}
%\newpage
% ------------------------------
\section{Introduction}
Recent empirical findings regarding financial time series have suggested that financial markets can be considered complex systems in which a large number of agents participate in trade and interplay with each other [1]. A goal of econophysics is to model the stochastic processes of financial markets as complex systems and to reproduce the statistical properties observed in the time evolution of financial prices by using tools of theoretical physics such as the master equation and the Fokker-Plank equation. Although most empirical studies so far have investigated the statistical properties of the time series of a single stock or index from theoretical point of view, it is more important to examine the statistical properties of ensembles of stocks. Lillo and Mantegna [2-4] studied the ensemble return distribution, and showed some statistical properties of the ensembles of stocks returns during the market's crash and rally days. 
In this paper we focus our attention on the statistical properties of ensembles of stock prices. Using $1200$ stocks traded in the Tokyo Stock Exchange, we formed ensembles of daily stock prices. 

\section{Power-law for ensembles of stock prices} 

Figure 1(a) shows a so-called Zipf-plot for an ensemble of stock prices of 1200 companies on May 1, 1989, on a double logarithmic scale, where $ S(n) $ denotes the stock prices in descending order ($ S(1) $ being the stock with the highest price, $ S(2) $ the stock with the second highest price, and so on). The Zipf-plot for stock prices higher than 800 yen shows a straight line and is well described by $ S(n) \sim n^{-\beta} $ with $ \beta = 0.46 $. Figure 1(b) shows the complementary cumulative distribution of the stock price ensemble on May 1, 1989. The tail of the distribution can be well described by a power-law decay, $ P(S > x) \sim x^{-\alpha} $ with $ \alpha = 2.17 $. The power-law exponent obtained by power-law regression fits the complementary cumulative distribution function, where the fit is for all $ x $ larger than $1000$\footnote{Note that the Zipf plot and cumulative plot are equivalent. The relationship, $ \beta = 1/\alpha $ holds theoretically. Thus the power-law exponent calculated from $ \beta $ is equal to $ 2.175 $. This value is very close to the power-law exponent estimated from the stock price ensemble's complementary cumulative distribution.}. 
To confirm the robustness of the above analysis, we repeated this analysis for each of the trading days in a 5-year period from January 4, 1988, to December 30, 1992, which corresponds to the periods of bubbles and crashes in the Japan stock market. They are all consistent with a power-law asymptotic behavior, and the power-law exponent $ \alpha $ moved in the range of $1.7 < \alpha < 2.2$. Figure 1(c) and Figure 1(d) show the movement of the power-law exponent $ \alpha $, and the movement of the mean value of the stock prices in the period from January 4, 1989, to December 30, 1992. If follows from the figures that the power-law exponent $ \alpha $ is positively correlated with the mean value of stock prices $ \bar{S} $. The correlation coefficient is equal to $0.88$. 

\section{Concluding remarks}

 These findings suggest that the financial markets are itinerating over a series of the non-equilibrium stationary states characterized by a power-law distribution. 
We found here that the power-law exponent $ \alpha $ is positively correlated with the stock price's mean value. This finding suggests that the stock market bubbles in the period from 1988 to 1990 are defined as the generally extraordinary rise of stock prices. In our recent work on the Japanese land market \cite{5}, we quantitatively investigated the statistical properties of an ensemble of {\it land prices} in Japan in the period from 1981 to 2002, corresponding to a period of bubbles and crashes. We found that the tail of the complementary cumulative distribution function of the ensemble in the high price range is well described by a power-law distribution, $ P(S>x) \sim x^{-\alpha} $, and the power-law exponent $ \alpha $ is negatively correlated with the mean value of land prices. Therefore, the real estate bubble is not defined as the generally extraordinary rise of land prices, but rather as an extraordinary enlargement of the inequality of land prices. Why did we obtain the empirically contradictory results regardless of the bubbles that occurred in the same period in Japan? The answer could be as follows. At the peak of the Japanese stock market in December 1989, Japanese stocks had a total market value of about 4 trillion dollars, almost 1.5 times the value of all U.S. equities and close to 45 percent of the world's equity market capitalization [6]. Thus, the Japanese stock market bubbles in the late of 1980s may be defined as the extraordinary rise of market capitalization of the Japanese stock market among the world's stock markets. This conjecture remains to be tested. 

\section{Acknowledgement}
Financial support by the Japan Society for the Promotion of Science under the Grant-in-Aid, No. 06632 is gratefully acknowledged. All remaining errors, of course, are ours.

\newpage
\begin{figure}
\begin{center}
  \includegraphics[height=18cm,width=14cm]{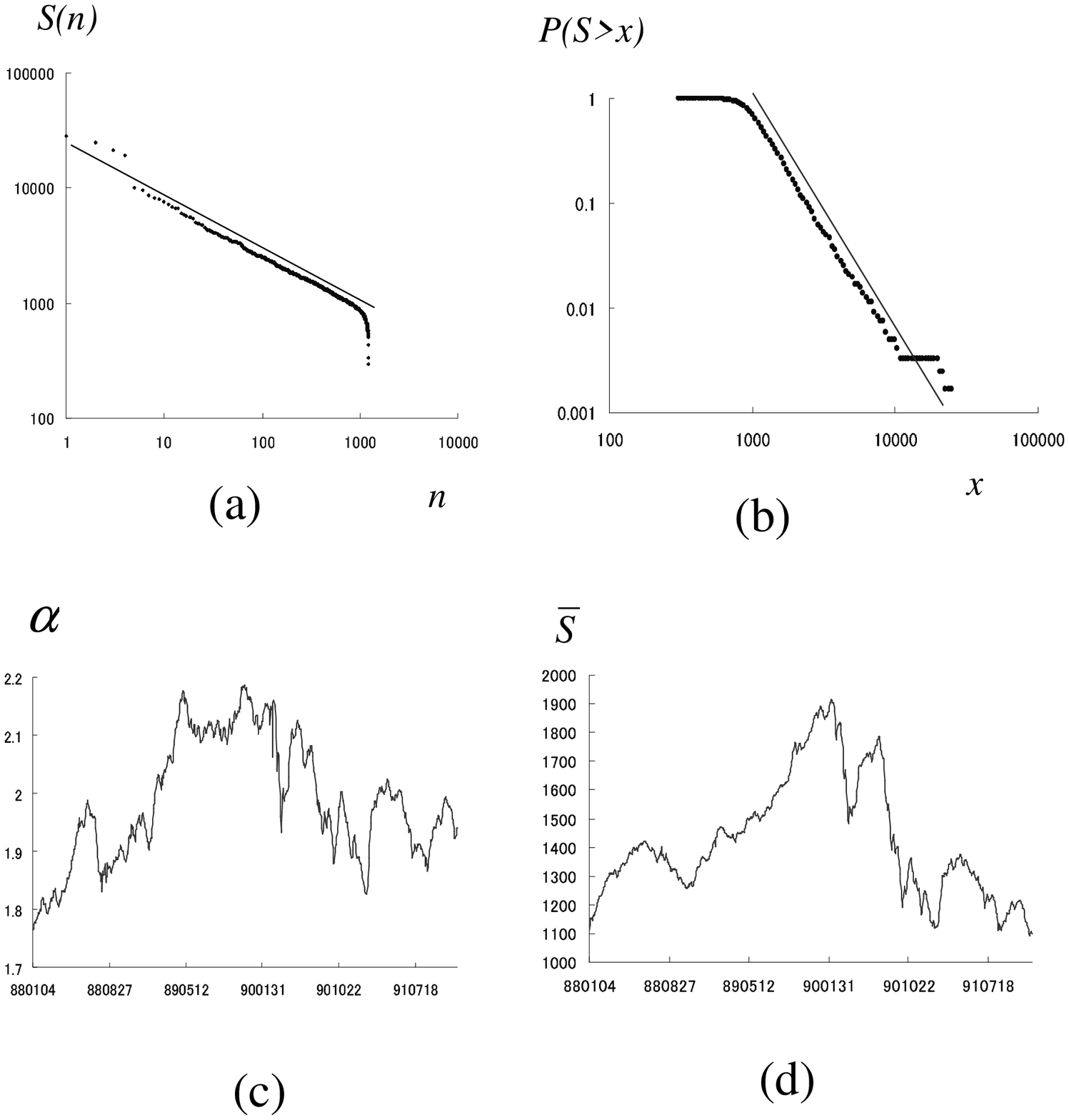}
\end{center}
\caption{(a) Zipf-plot for daily stock prices. $S(n)$ is the stock prices, $n$ is the descending rank. (b) Cumulative distributions $ P(S > x) $ of the ensemble of stock prices. We select the 1200 stocks traded in the Tokyo Stock Exchange and we form an ensemble of daily stock prices on May 1, 1989. (c) The movement of the power-law exponent $ \alpha $ in the 5-year period from January 1988 to December 1992. (d) The movement of the mean value of the stock prices $ \bar{S} $ in the 5 year period from January 1988 to December 1992.}
\label{fig1}
\end{figure}

\end{document}